\documentclass[manuscript]{acmart}

\AtBeginDocument{%
  }

\usepackage{graphicx}
\usepackage{enumitem}
\usepackage{xcolor}
\usepackage{amsmath}
\usepackage{booktabs}
\usepackage{stmaryrd}
\usepackage{tabularx}
\usepackage{multirow}
\usepackage{color,soul}
\usepackage{rotating}
\usepackage{subcaption}  

\newcommand{\methodName}{QualitEye}

\setcopyright{acmlicensed}
\copyrightyear{2018}
\acmYear{2018}
\acmDOI{XXXXXXX.XXXXXXX}

\acmJournal{JACM}
\acmVolume{37}
\acmNumber{4}
\acmArticle{111}
\acmMonth{8}

\begin{document}

\title{QualitEye: Public and Privacy-preserving Gaze Data Quality Verification}

\author{Mayar Elfares}
\email{mayar.elfares@vis.uni-stuttgart.de}
\orcid{0000-0002-6490-9139}
\affiliation{%
  \institution{Institute of Information Security, Institute of Visualisation and Interactive Systems, University of Stuttgart}
  \city{Stuttgart}
  \country{Germany}
}

\author{Pascal Reisert}
\affiliation{%
  \institution{Institute of Information Security, University of Stuttgart}
  \city{Stuttgart}
  \country{Germany}}
\email{pascal.reisert@sec.uni-stuttgart.de}

\author{Ralf Küsters}
\affiliation{%
  \institution{Institute of Information Security, University of Stuttgart}
  \city{Stuttgart}
  \country{Germany}}
\email{ralf.kuesters@sec.uni-stuttgart.de}

\author{Andreas Bulling}
\orcid{0000-0001-6317-7303}
\affiliation{%
  \institution{Institute of Visualisation and Interactive Systems, University of Stuttgart}
  \city{Stuttgart}
  \country{Germany}}
\email{andreas.bulling@vis.uni-stuttgart.de}

\renewcommand{\shortauthors}{Elfares et al.}

\begin{abstract}

Gaze-based applications are increasingly advancing with the availability of large datasets but ensuring data quality presents a substantial challenge when collecting data at scale. It further requires different parties to collaborate, therefore, privacy concerns arise.
We propose \methodName---the first method for verifying image-based gaze data quality.
\methodName{} employs a new semantic representation of eye images that contains the information required for verification while excluding irrelevant information for better domain adaptation.
\methodName{} covers a public setting where parties can freely exchange data and a privacy-preserving setting where parties cannot reveal their raw data nor derive gaze features/labels of others with
adapted private set intersection protocols.
We evaluate \methodName{} on the MPIIFaceGaze and GazeCapture datasets and achieve a high verification performance (with a small overhead in runtime for privacy-preserving versions). 
Hence, \methodName~ paves the way for new gaze analysis methods at the intersection of machine learning, human-computer interaction, and cryptography.
\end{abstract}

\ccsdesc[500]{Human-centered computing~HCI theory, concepts and models}

\keywords{Gaze, quality, privacy, private set intersection}


\maketitle

\section{Introduction} \label{sec:intro}
\begin{figure}[t]
\centering
\includegraphics[width=0.85\textwidth]{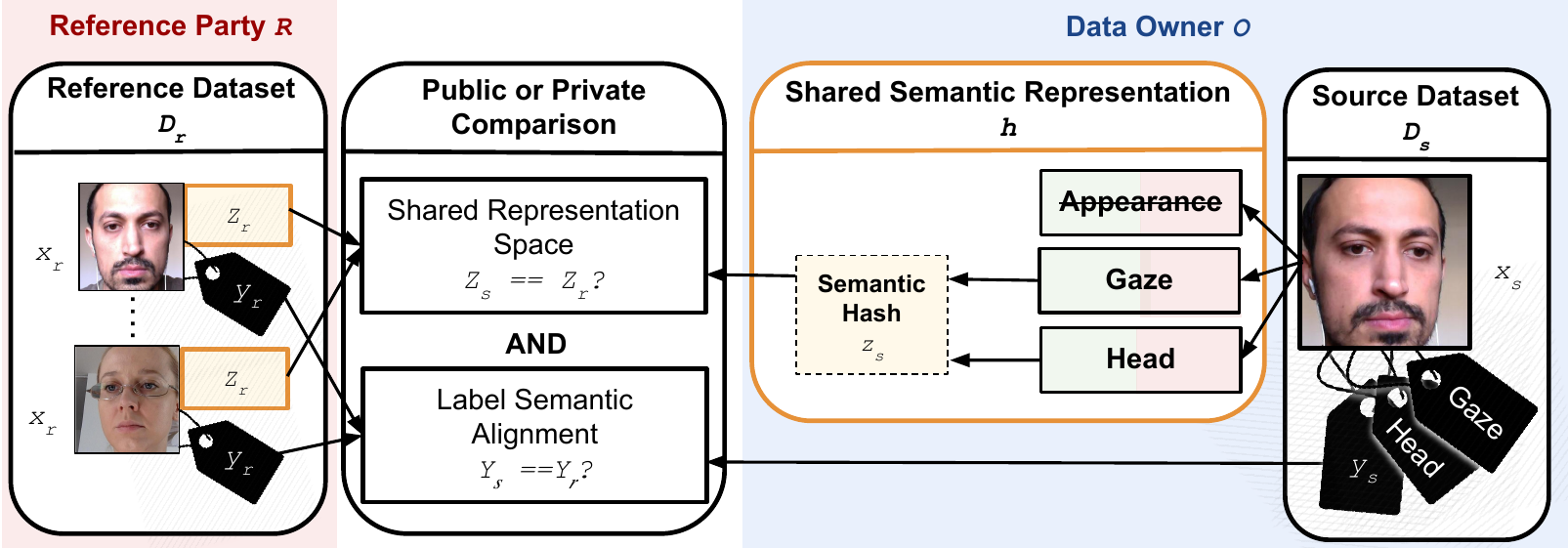}
\caption{
To verify the image-based gaze data quality,
source data owners $O$ (blue) and the reference party $R$ (e.g. a reliable source, red), first, disentangle the gaze direction and head pose features that correspond to the data labels, ignoring the cross-party irrelevant features (e.g. appearance) for a high domain adaptation performance instead of the raw pixel-wise data comparison. Then, features are semantically hashed for an efficient bit-wise comparison to obtain data-independent (i.e. not domain-specific), deterministic (i.e. produces the same outputs for similar semantics), and generative (i.e. learns the gaze data distribution) representations $z$ (shown in orange).
Then, they compare the hash values and corresponding labels against each other. to find the (mis)matching data samples.
C.f. Figure \ref{fig:VAE} for the shared semantic representation and Figures \ref{fig:DH} and \ref{fig:OPRF} for the comparison.}
\label{fig:main}
\end{figure}

Eye tracking has seen widespread adoption for numerous applications, such as for gaze-based human-computer interaction \cite{drewes2010eye,majaranta2014eye, shimata2015study, khamis18_imwut, khamis17_imwut}, for understanding the human visual system \cite{barr2008analyzing, abdelrahman19_imwut}, measuring user experience \cite{cowley2016psychophysiology, yasmine2025gazecopilot} or for computational user modelling \cite{bulling2011feature, abdou2022gaze, coutrot2018scanpath}.
With eye tracking becoming pervasive \cite{bulling2010toward} and increasingly integrated into personal devices \cite{huang2016building, huang2017screen, bace2020quantification}, recent years have also seen a significant increase in the availability of large-scale gaze datasets \cite{zhang15appearance, zhang19_pami, zhang2020eth, smith2013gaze, sugano2014learning}.
Traditionally, these datasets have been collected in research contexts, but are now increasingly collected and shared by private individuals and commercial enterprises
\cite{elfares2022federated, elfares24privateeyes}.

A challenge amplified by these advances that has largely been neglected in the gaze community so far is verifying the quality of the acquired, collected, or shared gaze data.
Although a few prior works \cite{holmqvist2012eye, deborah24reporting,feit2017toward, aziz2022assessment, lohr2019evaluating, aziz2024evaluation} focused on gaze data quality (e.g., evaluating the eye-tracking systems' accuracy, signal-to-noise ratio, or robustness), the verification aspect of the acquired gaze data is largely neglected, especially for image-based gaze data (e.g. features or labels).
In this work, we verify the quality of the eye images and their compliance with the respective labels (e.g. gaze direction and head pose) while ignoring irrelevant cross-user features (e.g. appearance). The quality verification ensures that similar eye features correspond to similar labels, by comparing the data samples with a relatively reliable source (e.g. a publicly available dataset or a trusted party). This is particularly important as gaze data quality can be subject to inaccurate labels or inconsistent features due to technical problems with the recording setup, choice of data preprocessing methods, or calibration and systematic errors.


Formally, as shown in Figure \ref{fig:main}, let the \emph{source dataset} be
$D_s = \{(x_s^{(i)}, y_s^{(i)})\}_{i=1}^{N_s}$,
where $x_s^{(i)} \in \mathcal{X}$ denotes an image (e.g., eye or face image) and
$y_s^{(i)} \in \mathcal{Y}$ denotes the associated gaze-related label (e.g., gaze direction or head pose).
The source dataset corresponds to the dataset whose gaze data quality is to be verified.
Let the \emph{reference dataset} be
$D_r = \{(x_r^{(j)}, y_r^{(j)})\}_{j=1}^{N_r}$,
which contains image samples and the same label semantics as the source dataset (e.g., head pose and gaze direction), is assumed to have trusted or known data quality, and is used as a baseline for comparison.
Therefore, we present \methodName, a computational framework for gaze data quality verification. We first propose a new shared semantic representation function
$h : \mathcal{X} \rightarrow \mathcal{Z}$,
which maps raw images to a latent semantic space $\mathcal{Z}$ that preserves gaze-relevant information while discarding subject-specific or appearance-related factors.
The corresponding semantic representations are
$z_s^{(i)} = h(x_s^{(i)}), \quad
z_r^{(j)} = h(x_r^{(j)})$.
The source dataset $D_s$ and reference dataset $D_r$ are said to be \emph{compatible} if the following conditions hold:
\begin{enumerate}[leftmargin = *]
    \item \textbf{Shared Representation Space:
    The representations of both datasets lie in the same semantic space, i.e., 
    $z_s^{(i)}, z_r^{(j)} \in \mathcal{Z} \quad \forall i,j$, 
    and $\mathcal{Z}$ encodes gaze-relevant semantics consistently across datasets.}

    \item \textbf{Label Semantic Alignment:}
    The label spaces $\mathcal{Y}_s$ and $\mathcal{Y}_r$ are identical or there exists a mapping  
    $\phi : \mathcal{Y}_s \rightarrow \mathcal{Y}_r$    
    such that $\phi(y_s^{(i)})$ is semantically comparable to $y_r^{(j)}$ (e.g., consistent gaze coordinate systems).

\end{enumerate}


The need for quality verification is further amplified when raw gaze data cannot be shared due to privacy constraints, necessitating privacy-preserving processing \cite{elfares2022federated, elfares24privateeyes, kandappu2018obfuscation}.
Accordingly, \methodName~supports:
\begin{enumerate}[leftmargin=*]
    \item \textbf{Public verification} for within-dataset consistency and comparison to public references.
    \item \textbf{Privacy-preserving verification} for remote comparison against private reference datasets without data disclosure. In this case, \methodName ensures that: (i) Raw eye images and sensitive features are never directly exchanged during verification, (ii) Only essential semantic information for quality assessment is used, which is disentangled from irrelevant or privacy-sensitive attributes, (iii) Cryptographic protocols prevent inference of another party’s underlying data beyond what is needed for quality comparison.
\end{enumerate}

To evaluate \methodName, we present appearance-based gaze estimation as our guiding example since it is the basic building block of gaze-based applications and has well-established publicly available datasets for evaluation. Nonetheless, \methodName~does not restrict how the dataset should be processed after the quality verification; hence, \methodName~is domain-, task-, and model-agnostic.
We thereby validate our method through extensive experiments on the well-established full-face appearance-based gaze estimation datasets, MPIIFaceGaze \cite{zhang19_pami} and GazeCapture \cite{krafka2016eye}, and achieved a gaze quality metric (Matthew Correlation Coefficient \cite{matthews1975comparison}\footnote{The Matthews Correlation Coefficient (MCC) is a metric used to assess the quality of binary classifications (i.e. match or mismatch) and it is particularly useful in cases where there is a class imbalance (e.g., when one class is much more frequent than the other), making it more reliable than metrics like accuracy in these situations (cf. Section \ref{sec:experiments}).}) of 0.92 and 0.94, respectively, with a negligible overhead in runtime for the privacy-preserving setups compared to other baselines.

\methodName, therefore, solves a fundamental concern-- gaze data quality--in many eye-tracking applications including: 
(i) \textbf{Data collection and cleaning}, enabling validation of newly acquired gaze data against trusted sources \cite{blass2022private}, 
(ii) \textbf{Auto-labelling}, where labels are inferred from limited local annotations or reliable external datasets \cite{sager2021survey},
(iii) \textbf{Remote learning and analytics}, e.g., federated learning \cite{elfares2022federated, elfares24privateeyes}, where verifying the quality of locally held private data is critical, 
(iv) \textbf{Try-before-you-buy services} \cite{song2021try}, allowing users to assess gaze data or model predictions prior to acquisition, and
(v) \textbf{Predictive benchmarks and leaderboards} (e.g., Kaggle \cite{Kaggle}), enabling private test-set verification by comparing labels without model disclosure.

In summary, this work makes the following contributions\footnote{Note that, 
\methodName~ does not assume that any single reference dataset is universally representative of all possible gaze data distributions; instead, it explicitly acknowledges and mitigates reference–source mismatch as a fundamental challenge. The framework formulates quality assessment as a problem of \textit{relative quality verification} rather than absolute performance prediction, thereby conditioning its conclusions on the characteristics of the chosen reference data (c.f. Section \ref{sec:experiments}). This design choice reflects a well-known limitation in computer vision, namely that universal ground truth distributions are generally intractable to obtain, particularly for eye and gaze imagery, where data are inherently non-independent and non-identically distributed due to variations in subject appearance, head pose, lighting conditions, capture devices, and annotation protocols \cite{elfares2022federated}.
Furthermore, \methodName~ is explicitly designed to support the use of multiple reference datasets, enabling cross-reference validation and reducing dependence on any single dataset’s coverage or bias. Consistent verification outcomes across diverse references strengthen confidence in the assessed quality, while discrepancies serve as indicators of domain mismatch or insufficient reference diversity.}:

\begin{itemize}[leftmargin=*]
    \item \methodName~is the first work to investigate the problem of image-based gaze data quality verification.
    
    \item Instead of the raw pixel-wise data comparison, we propose a new generic hashed representation learning model that disentangles the gaze-specific features and ignores the cross-users' irrelevant features (e.g. appearance).

     \item We propose methods for public and privacy-preserving gaze data quality verification. For the latter, we extend existing privacy-preserving protocols with semantic similarities and label matching to handle the different privacy requirements.
     
\end{itemize}


\section{Related Work} \label{sec:related_work}

In the following, we summarise previous work that is most closely related to our method for gaze data quality verification.
\methodName~ focuses on the verification of (i) gaze data quality through (ii) unsupervised gaze representation learning to compare relevant features for both public and (iii) private gaze data verification.

\paragraph{Gaze data quality} 
The lack of large-scale, diverse datasets remains a key limitation in eye-tracking research, hindering the study of variability across users, tasks, and settings \cite{gressel23_pcm, elfares2022federated}. Despite advances in gaze data acquisition and processing, quality standards necessary for high-performing models are often neglected \cite{holmqvist2012quality}, primarily due to (i) the time-intensive collection and labeling of large datasets, (ii) data owners’ reluctance to share private eye data, and (iii) the difficulty for single entities to gather diverse data at scale \cite{steil2019privacy}. Recent work has emphasised standardised gaze data quality reporting, including metadata such as sampling rate, tracked eye(s), filter settings, recording duration, and display resolution, to ensure accessible, reusable, and reproducible datasets \cite{deborah24reporting}. While a few studies have assessed eye-tracking system accuracy and signal-to-noise ratio \cite{feit2017toward, aziz2022assessment, lohr2019evaluating, aziz2024evaluation}, we propose automatic quality verification for appearance-based gaze data without manual reporting, applicable both publicly and in a privacy-preserving manner.

\paragraph{Unsupervised gaze representation learning}
Supervised representation learning has been used to extract task-specific eye features, such as gaze estimation \cite{zhang15appearance, zhang2020eth} or eye contact detection \cite{zhang2017everyday}, but it relies on carefully labelled data and does not generalise well across tasks. In contrast, self-supervised learning (SSL) enables models to learn underlying image features directly from the data without task-specific labels, capturing subtle, person-specific variations in appearance-based gaze methods \cite{goyal2021self, elfares2022federated}. Early SSL approaches used joint embedding architectures (e.g., siamese networks) \cite{liu2018differential, sun2021cross, yu2020unsupervised}, but these often collapsed, producing identical embeddings. Contrastive methods address this by learning from positive and negative pairs, yet enumerating all possible pairs is intractable, leading to bias in hand-selected examples \cite{jindal2023contrastive}. Non-contrastive approaches \cite{caron2019unsupervised, yan2020clusterfit, chen2021exploring} theoretically require optimising latent capacity, which is often intractable. Here, we propose a VAE-based approach for gaze representation learning that is generative, non-contrastive, and leverages a ''fuzzy'' latent variable, with a neural network providing amortised optimisation across gaze data points.

\paragraph{Privacy-sensitive gaze data} 
Gaze data can contain personal identifiers \cite{yang2022wink, cantoni2015gant, sammaknejad2017gender}, sensitive attributes \cite{steil2015discovery, huang2016stress, yaneva2018detecting, hoppe2018eye}, or business-related information (e.g., devices or participant details \cite{bozkir2023eye}) that cannot be shared. Privacy threats in eye tracking have been largely underexplored, partly because traditional cryptographic and privacy-preserving techniques (PPTs) \cite{archer2023handbook} were considered too computationally expensive. Recent advances, however, make approaches such as federated learning (FL) \cite{elfares2022federated}, differential privacy (DP) \cite{steil2019privacy, li2021kaleido, bozkir2021diffrential, liu2019differential}, and secure multi-party computation (MPC) \cite{elfares24privateeyes} practical for gaze-based tasks. \methodName~is the first work to address gaze data quality verification while explicitly incorporating privacy considerations.

\paragraph{Privacy-preserving techniques}
Secure data comparison techniques \cite{bloemen2024large, uzun2021fuzzy, mouris2024delegated, adir2022privacy, MicrosoftPhotoDNA, della2024trust, AppleNeuralhash, MetaPDQ, li2023therapypal} enable similarity detection between datasets while preserving privacy. Traditional methods use cryptographic tools such as homomorphic encryption (HE) \cite{gentry2009fully}, secure multiparty computation (MPC) \cite{yao1982protocols, goldreich2019play}, and oblivious transfer \cite{rabin2005exchange}. More recent approaches employ private set intersection (PSI) protocols \cite{de2012experimenting, bellare2003one, camenisch2009private, camenisch2009blind}, which reveal only the intersection (or its size), operate on a single party’s input, and exchange cryptographic hashes, reducing computation and communication overhead. These techniques are used in biometric verification \cite{bloemen2024large, garcia2024pass, contreras2022dev, uzun2021fuzzy}, data linkage \cite{mouris2024delegated, duong2020catalic, adir2022privacy, MicrosoftPhotoDNA}, fraud detection \cite{zhong2024oryx, della2024trust, AppleNeuralhash, MetaPDQ}, and genomics \cite{kim2015private, almadhoun2020differential, li2023therapypal}, typically with tolerance for errors from formatting, typos, or biological variations. In contrast, gaze-based applications are highly sensitive to subtle variations in gaze and head pose, requiring a novel privacy-preserving comparison that focuses only on task-relevant features, excludes appearance, minimizes data transfer, and reduces the risk of information leakage.

\section{\methodName} \label{sec:method}


Given our setup and formal definitions in Section \ref{sec:intro}, in this paper, we present a most common leading gaze example - appearance-based gaze estimation with eye images and their corresponding gaze direction and head pose labels\footnote{Our approach is not limited to gaze estimation and can be extended to other gaze-based applications. Nonetheless, we choose gaze estimation as our guiding example since it is the basic building block of gaze-based applications and has well-established publicly available datasets for evaluation.}. 
We assume a horizontal data distribution where each party's dataset includes different data samples, e.g. data of different participants and different numbers of samples. 
The data owner wishes to verify the quality of the gaze dataset with respect to the reference dataset to overcome the data collection and labelling problems mentioned above. 

We define the gaze dataset quality as the number of data samples (the image features and the corresponding labels) that comply with the reference dataset, i.e. the cardinality of the intersection set. 
More formally, the set of mismatching data samples would be $\{(x_i^O, y_i^O)| \exists (x_i^R, y_i^R) : x_i^O = x_i^R \land  y_i^O \neq y_i^R\}$. 

\paragraph{\textbf{Overview}} 
In both the public and private evaluation setups, as shown in Figure \ref{fig:main},
\methodName~ operates by first mapping samples from the source and reference datasets into a shared semantic representation space using the representation function described in Section \ref{subsec:semantic}. 
In the public setup (\methodName-$V_0$), semantic representations of the reference dataset are directly available, allowing quality verification to be performed through explicit similarity and matching operations. 
In the private setup, where raw data and representations cannot be shared due to privacy or proprietary constraints, \methodName~ supports multiple alternative protocols (\methodName-$V_1$ to $V_4$) that provide different trade-offs between privacy guarantees, computational requirements, and verification fidelity, allowing \methodName~ to adapt to a wide range of real-world deployment scenarios while maintaining its core objective of relative data quality verification.

\subsection{Semantic gaze representations}\label{subsec:semantic}

Since the direct pixel-wise comparisons of the gaze images only capture exact matches. To obtain more meaningful results, we increase the efficiency of comparisons by first encoding the images in a semantic representation.
Our goal is to learn the semantic representations of the data samples that reflect the similarity of the gaze-based information (e.g. gaze direction and head pose) while ignoring other cross-party irrelevant features (e.g. appearance). In addition, this representation should be deterministic (i.e. produces the same outputs for similar semantics at different parties), generative (i.e. learns the gaze data distribution), and domain-agnostic (i.e. not dataset-specific) to be able to generalize well to different (unseen) datasets at different parties. 

\begin{figure}
\centering
\includegraphics[width=0.8\textwidth]{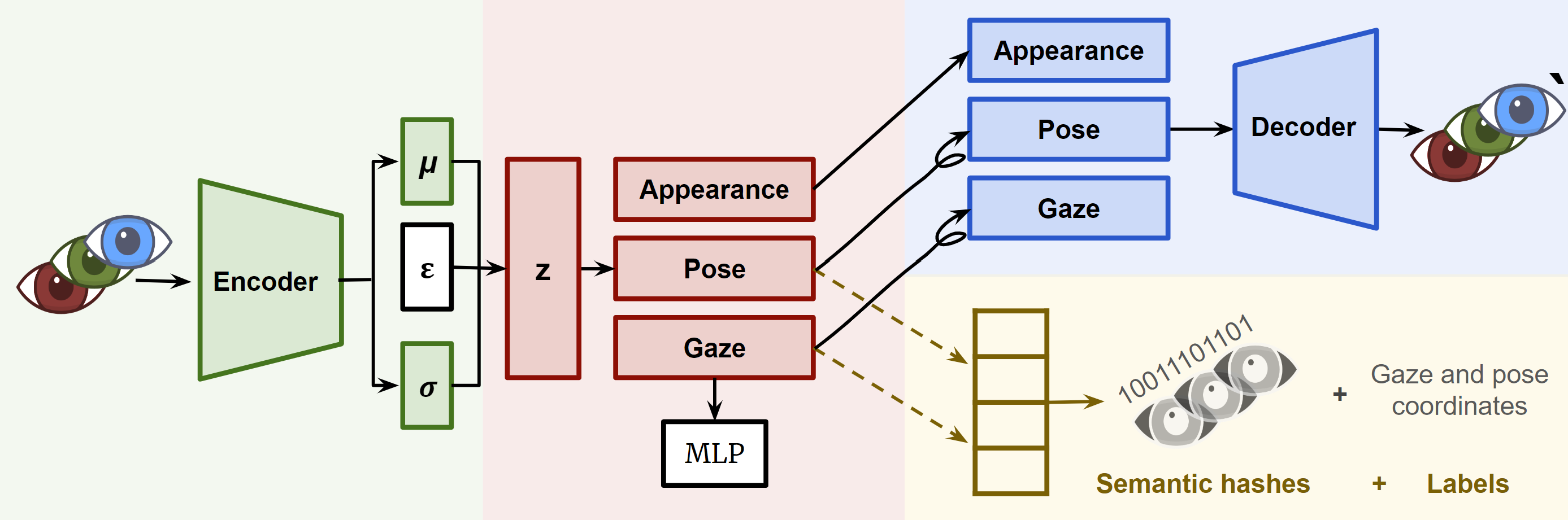}
\caption{To obtain the hashed semantic representations used for label-based data comparison, \textit{1- green:} eye images are encoded into a latent vector $z$. \textit{2- red:} Then, the gaze and head pose information are disentangled from the appearance via some transformations and a multi-layer perception (MLP). \textit{3- blue:} During training, the appearance, gaze, and pose are passed as a latent vector to a decoder that reconstructs the transformed eye images. \textit{4- yellow:} Finally, the gaze and head pose latent codes are hashed and are passed as inputs to the PSI protocol along with the corresponding labels.}
\label{fig:VAE}
\end{figure}

\paragraph{Variational auto-encoder (VAE)} As shown in \autoref{fig:VAE}, our VAE is composed of four main components:
\begin{itemize}[leftmargin = *]
    \item \textbf{Top-down encoder $E_\phi$} that maps the input $x$ to a latent $z$ corresponding to a variational distribution 
    and outputs the parameters of the distribution (e.g. mean $\mu$ and variance $\sigma$). In our case, we use a multivariate Gaussian distribution $N(x|\mu, \sigma)$ due to the nature of RGB eye images and to estimate the average of the gaze data distribution (with likelihood $p_\theta (x|z)$) according to the \textit{central limit theorem} for a better cross-domain adaptation. Then the prior $p_\theta (x)$ is a mixture of Gaussian distributions, and the posterior distribution $p_\theta (z|x)$ can be approximated to $q_\phi (z|x)$ (i.e. the amortized optimization).
    
    \item \textbf{Bottom-up decoder $D_\theta$} maps the latent to the input space. Hence, our problem is to compute the conditional likelihood distribution $p_\theta (x|z)$ by the probabilistic decoder and the approximated posterior distribution $q_\phi(z|x)$ by the probabilistic encoder.
    
    \item \textbf{Feature disentanglement:} For a better domain adaptation, our model explicitly learns to disentangle the gaze direction and head pose representations as equivalent input rotations. This is achieved by training the model in a person-independent manner \cite{liu2018differential} and splitting $z$ into three sub-vectors, (i) gaze direction, (ii) head pose, and (iii) appearance (more specifically, all other information found in the image), similar to \cite{park2019few}, and rotating the latent sub-vectors using rotation matrices to the frontal angle and then to certain yaw and pitch angles. That is, for the same person, the model learns to transform the gaze direction and head pose of one image into the other by multiplying the sub-codes by a rotation matrix and optimizing a pixel-wise $L1$ loss function over the entire encoder-decoder and an MLP regression for the gaze sub-code.

    The aim of disentangling gaze direction and head pose from the rest of the image is to (i) only compare the information that is relevant to the ground-truth labels and (ii) to minimize the amount of information exchanged in the cross-party setup (i.e. data minimization) for better privacy, runtime, and communication.

    \item \textbf{Hashing:} Once the latent vector is disentangled, the gaze direction and head pose sub-codes are hashed with a locality-sensitive hash function.  Locality-sensitive hashing (LSH) is a fuzzy hashing technique that maps similar inputs to the same hash value with a certain probability. 
    Such hashes (i) reduce the dimensionality of the semantic representations, which likelyimproves efficiency (e.g. computational runtime and communication), (ii) allow efficient comparisons of the bit-wise representations (i.e. comparing the hashed values in the hamming space instead of the latent space), (iii) are data-independent (i.e. not domain-specific to generate the same hash for different inputs at different parties), and most importantly, (iv) produce identical hashes for images with similar features (i.e. to tolerate small systematic errors in data acquisition commonly found in eye tracking data).
\end{itemize}

VAEs are specifically interesting for gaze-based data because of (i) the inherent mutual information in appearance-based eye data (higher mutual information yields better disentanglement \cite{chen2016infogan}), (ii) the independence between the latent variables (e.g. gaze direction and head pose) encourages interpretability yielding better semantics \cite{chen2016infogan, higgins2017beta}, and (iii) they are more generalisable\footnote{The generalisation capability makes VAEs more robust against adversarial attacks \cite{alemi2016deep}. This increases privacy but remains out of the scope of this paper.} \cite{alemi2016deep}.

\section{Comparing Hashes in the Public Setting (\methodName-V$_0$)}
As the raw pixel-wise eye image comparison cannot be used for quality verification given the cross-user variations in appearance, the hashed disentangled gaze direction and head pose representation can be used instead. Hence, a gaze data owner $O$ can verify the within-dataset consistency by checking that similar representations of gaze direction or head pose have similar respective labels. Additionally, in case of systematic or calibration errors (e.g. resulting from changes in user position during data collection), $O$ can compare the collected dataset against the publicly available datasets (as the reference) to either check for possible errors (i.e. correcting the non-compliant data samples) or for auto-labelling the dataset. 

However, in other scenarios, the datasets might not be available at one party, therefore, $O$ might need to verify the dataset quality against a different dataset owned by another reference party $R$. A (public) solution would be for one party (e.g. $R$) to send their data as hashed disentangled representations along with the labels to the other party (e.g. $O$) for comparison.

\section{Comparing Hashes in the Private Setting (\methodName-$V_1$ to $V_4$)} \label{sec:private}

The (public) solution mentioned above does not guarantee privacy as $O$ can perform a dictionary attack (i.e. tries all possible hashes of the input space) and recover the plain representations, especially when the dictionary has a computationally reasonable size, e.g. in the case of gaze estimation. Note that, even if the plain representations do not include the appearance and the raw images cannot be reconstructed \cite{elfares24privateeyes} (via our data minimization step of disentanglement), $O$ can still deduce information beyond the (mis)matching samples such as the number of samples in $R$'s dataset, the semantic meaning of all other samples, the plaintext labels, the kind of error (if any)... etc.
Therefore, a better solution is to use a cryptographic solution with formal provable guarantees -- private set intersection (PSI), where both parties interactively compute the intersection (i.e. one party cannot compute the intersection without the other party's help, e.g. mitigating dictionary attacks).
However, since PSI usually come with a computational overhead or a drop in utility, we, therefore, present several versions of \methodName~ with different tradeoffs between efficiency (i.e. runtime and communication), privacy, and utility. 

We assume a semi-honest (a.k.a. honest-but-curious) security model, i.e. parties will not deviate from the defined protocol; however, they might try to learn possible information from the legitimately received messages. Note that, in this work, an adversary refers to the main parties $R$ and $O$.



\begin{figure*}
\centering
\includegraphics[width=0.7\textwidth]{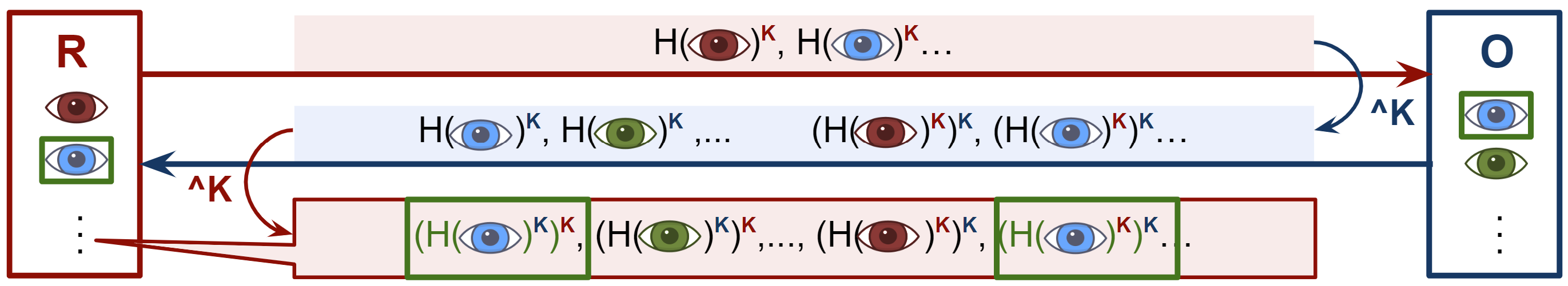}
\caption{In \textbf{\methodName-V$_{1}$}, both parties $O$ and $R$ exchange their elements as hashed values raised to their private keys. Then, they raise the other party's received values again to their private keys. $R$ then start the comparison to find the matching inputs. Note that, we further send the eye labels (e.g. gaze direction and head pose) as an additional encrypted payload to each element. In \textbf{\methodName-V$_{2}$}, $O$ shuffles the second message to only reveal the cardinality of the intersection. In \textbf{\methodName-V$_{3}$}, $R$ includes the first message, e.g. as a package, before the start of the protocol. In all versions, only the private keys are secret, and all other values are sent in the clear. Security is still guaranteed due to the hardness of
the 'discrete logarithm problem', i.e. it is hard to infer the private keys \cite{diffie1976new}.} 
\label{fig:DH}
\end{figure*}

\begin{figure*}
\centering
\includegraphics[width=0.5\textwidth]{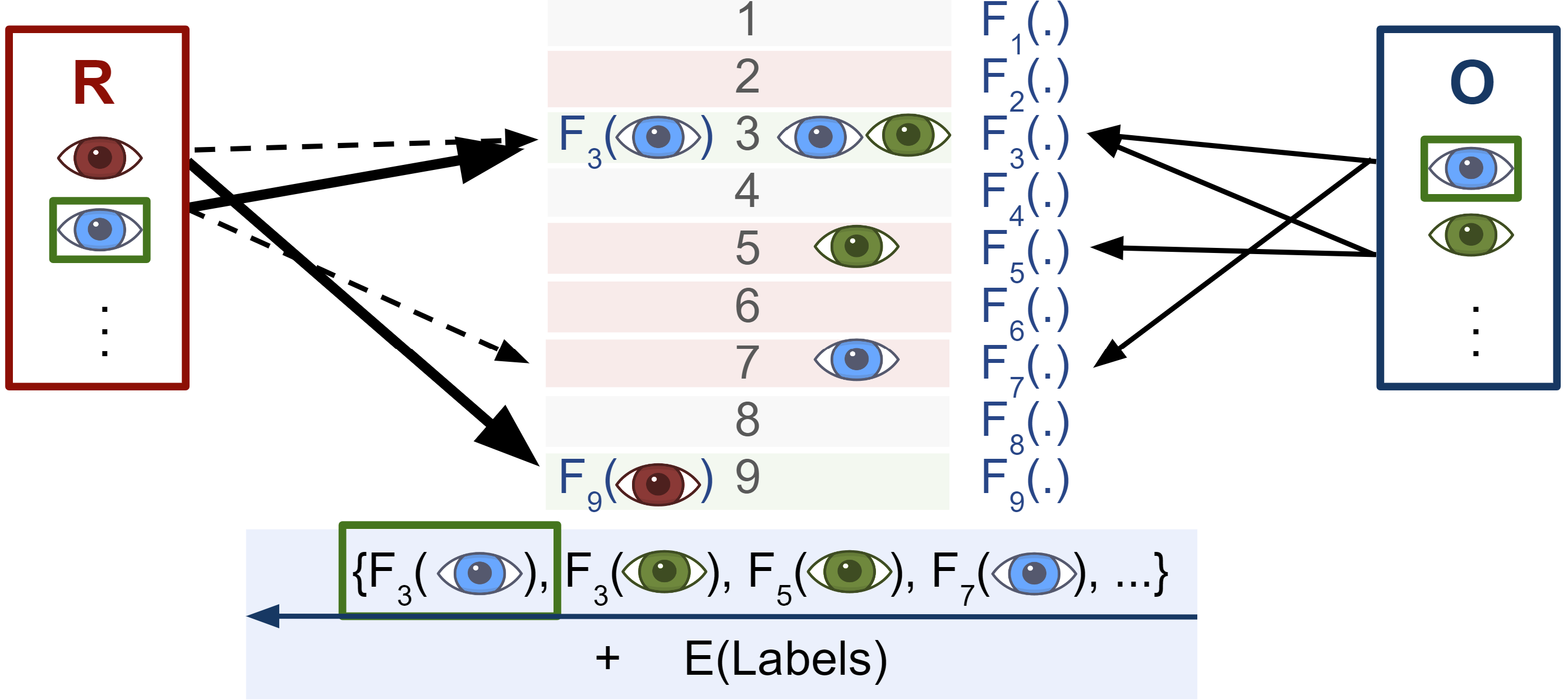}
\caption{In \textbf{\methodName-V$_{4}$}, each party computes two values per element (i.e. the PRF output of the hashed semantic representations). $R$ only selects one value per element (via cuckoo hashing). $O$ send his values (along with the encrypted blinded labels) to $R$. $R$ sends back all received information along with her encrypted blinded labels to $O$, who makes the comparison and finds the (mis)matching elements.
}
\label{fig:OPRF}
\end{figure*}

\paragraph{\textbf{Private set intersection (PSI) Preliminaries}}
PSI is a secure multiparty computation (MPC) protocol that allows two (or more) parties to compare elements in their sets by computing the intersection without revealing any information beyond this intersection. Recently, PSI constructs \cite{de2012experimenting, bellare2003one, camenisch2009private, camenisch2009blind} 
included different adversarial models, efficiency tradeoffs, and security guarantees. In this paper, we focus on the semi-honest (a.k.a. honest-but-curious) adversarial model \cite{goldreich2004foundations}.
This assumes that different data owners have a mutual interest in working together and improving the quality of their own datasets. We therefore assume that they stick to the rules and follow a previously agreed protocol. However, they nevertheless are happy to gain any information leaked by the protocol. Such data owners are called honest-but-curious. \methodName~ uses different cryptographic primitives to ensure that a curious data owner gains only minimal knowledge about other parties' data, e.g. oblivious transfer.
PSI constructions can include different cryptographic primitives. \methodName~relies on the following primitives:
\begin{itemize}[leftmargin = *]
    \item \textbf{Key agreement protocols:} In cryptography, key agreement protocols allow two or more parties to agree on a cryptographic key. Among these protocols, the Diffie–Hellman protocol \cite{diffie1976new} is one of the earliest practical protocols.
    More specifically, we base our privacy-preserving constructions of \methodName$_{v1, v12, v3}$ on a Diffie-Hellman-based PSI protocol where,  as shown in \autoref{fig:DH}, the two parties 
    $O$ and $R$ hash all their data samples $x_i^j$ \footnote{Parties get a cryptographic hash of the semantic hash values using their private keys $K_O$ and $K_R$. Samples are hashed to a primitive root modulo $p$ where $p$ is a large prime number that parties agree on, i.e. $H(x_i^j)^{K_j}\mod p$. In modular arithmetic, a number $g$ is a primitive root modulo $n$ if every number $a$ coprime to $n$ is congruent to a power of $g$ modulo $n$. That is, $g$ is a primitive root modulo $n$ if for every integer $a$ coprime to $n$, there exists some integer $k$ for which $g^k \equiv a (mod n)$. In other words, every invertible number is of the form $g^k$ for some integer $k$.},
    and raise the hashed values to their private keys.
    Then, these values are exchanged and compared by $O$.
    A match means that both $x_i^O$ and $x_i^R$ are similar. A mismatch reveals no information about the inputs\footnote{
    As proved by Diffie-Hellman \cite{diffie1976new}, raising $H(x_i)$ to an exponent (i.e. the secret key) makes it indistinguishable from random (i.e. the protocol hides the inputs when there is no match), even if the exponent is used elsewhere (i.e. it is safe to reuse $H(x_i)^{K_j}$ as in \methodName$_{v3}$).}. 
    This way, dictionary attacks are not possible since $R$ is committed to certain values in the first message, and the intersection requires the interaction between both parties to get both private keys (i.e. $O$ will not remain available for $R$ to try all different dictionary entries). Note that an eavesdropper can intercept the communication but does not have access to the private keys; therefore, the eavesdropper cannot infer the intersection \cite{huberman1999enhancing}.

    \item \textbf{Oblivious transfer (extension):} Oblivious transfer (OT) \cite{rabin2005exchange} and Oblivious transfer extension (OTe) \cite{beaver1996correlated, ishai2003extending} are cryptographic protocols where one party sends one of many pieces of information to a receiver, but remains oblivious as to what exact piece has been sent.

    \item \textbf{Oblivious pseudorandom functions (OPRF):} 
    OPRF is a cryptographic protocol where two parties jointly compute a pseudorandom function (PRF), i.e. a function which emulates a random oracle.
    Similar to OT, the sender does not learn any information about the other party's input, i.e. the sender is oblivious to what exact values has been sent.
    More specifically, we base our PSI construction of \methodName$_{v4}$ on an OPRF-based construction \cite{kolesnikov2016efficient}. 
    As shown in \autoref{fig:OPRF}, both parties hash their inputs to two hash values. $R$ only selects one of the two hash values (using cuckoo hashing \cite{pagh2001cuckoo}). $O$ sends all his inputs (as two PRF outputs per input corresponding to the two hash values) to $R$. $R$ compares these outputs to his to compute the intersection. This way, $R$ only learns the matching elements while all other elements in $O$'s dataset look random to him, and $O$ learns nothing about $R$'s inputs.
    
\end{itemize}

\paragraph{\textbf{Adaptation of existing PSI protocols}}

PSI protocols typically operate on input messages to find matches between two (or more) sets. In our case, three additional challenges arise:
\begin{enumerate}[leftmargin = *]
    \item The comparison of the raw eye images only reflects the exact similarity. We solve this issue by comparing the hashed semantic representations of the inputs in each dataset (generated in \autoref{subsec:semantic}).
    \item In addition to finding matching elements in the parties' sets, we further need to check the compliance of the labels. In the following, we solve this problem by extending the protocols with additional payloads.
    \item The accuracy of some labels (e.g. gaze direction) in the reference gaze-based data can include some error, currently >3 degrees for gaze direction \cite{wang2023investigation}.
    Therefore, we allow some error tolerance for the semantic representations and their respective payloads.
\end{enumerate}

\paragraph{\textbf{\methodName~ and the Different Privacy Tradeoffs}} 
\methodName~ adapts privacy-preserving gaze data verification  according to different efficiency tradeoffs. The different versions cover:
(i) Dataset sizes: hundreds vs thousands of samples, (ii) difference in datasets size: symmetric (i.e. when both parties have the same amount of data) vs asymmetric (i.e. when one party has a relatively larger dataset) dataset distributions, (iii) resulting information: the intersection set vs its cardinality, (iv) receiver of this information: one vs both parties.

\begin{itemize}[leftmargin=*]
    \item \textbf{\methodName-V$_{1}$} can be used when both parties have relatively small datasets (i.e. tens to hundreds of samples) and would like to know the exact (mis)matching samples. More specifically, \methodName-V$_{1}$ is based on the Diffie-Hellman key exchange protocol, as shown in \autoref{fig:DH}.
    
    \item \textbf{\methodName-V$_{2}$} can be used when both parties have relatively small datasets (i.e. tens to hundreds of samples) but are only interested in knowing the cardinality of the intersection (i.e. the size of the (mis)matching samples and not the exact samples). Similarly, \methodName-V$_{1}$ can be extended to only reveal the cardinality of the intersection (i.e. the intersection size and not the exact samples). In \methodName-V$_{1}$, $R$ sends $M2b$ (\autoref{fig:DH}) in the same order sent by $O$ so that $O$ can find the corresponding elements in her set. In \methodName-V$_{2}$, $R$ shuffles those elements before sending them to $O$. Hence, $O$ can only count the number of matching elements but cannot map them to the raw elements (i.e. elements appear uniform).
    
    \item \textbf{\methodName-V$_{3}$} can be used when one party has a relatively small dataset while the reference party owns a larger dataset (i.e. thousands to billions of samples). In this case, $M1$ 
    in \methodName-V$_{1}$ and \methodName-V$_{2}$ (\autoref{fig:DH}) can be sent in advance (e.g. offline as a built-in package in an eye-tracking software or as a publicly-published data) as it does not depend on the other party's input. Note that this does not break privacy and can be shared with multiple parties (or protocol instances) \footnote{Kales et al. \cite{kales2019mobile} further proposed an efficient encoding mechanism that could be used with $M1$ to enhance efficiency.}.

    \item \textbf{\methodName-V$_{4}$} can be used when one or both parties own large datasets where the previous versions are highly inefficient (c.f. \ref{sec:experiments}).
    As shown in \autoref{fig:OPRF}, \methodName-V$_{4}$ is based on a OPRF-based PSI protocol \cite{kolesnikov2016efficient}. \methodName-V$_{4}$ can be an alternative to \methodName-V$_{3}$ when changes to the data of the reference party are frequently made.
\end{itemize}

For all versions, the final information can be revealed to (i) one party, depending on which party starts the protocol or (ii) both parties with an additional communication step containing the resulting information (i.e. the intersection set or its cardinality).
We further propose the following adaptations:
In addition to the hashed semantic representations of the gaze direction and head pose, both parties input the corresponding labels encrypted as Elgamal ciphertext (an asymmetric key encryption based on the Diffie-Hellman key exchange) under their private keys. For instance, in \autoref{fig:OPRF}, $O$ sends the PRF outputs along with the corresponding encrypted (under $O$'s key) labels. If $R$ finds a matching PRF representation in his dataset, he encrypts his labels and sends both encrypted labels to $O$. If there is no match, $R$ re-encrypts $O$'s label. Then, $O$ decrypts both labels to find the mismatching inputs. Note that $O$ cannot distinguish between the cases where a matching representation does not exist in $R$'s dataset and a matching representation exists with a matching label; he only learns the mis-matched labels (i.e. the non-compliant samples). On the other hand, $O$ only learns the cardinality of the matching set. Additionally, parties do not learn the labels in the clear as labels are further blinded following \cite{blass2022private}.

Furthermore, to accommodate the (unavoidable) systematic errors in the labels (e.g. 3 degrees for gaze direction) in the hashing step, similar to the gaze representations, we adjust the probability that different latent codes are mapped to the same hash values and drop the least significant bits in the labels accordingly. The parties agree on the exact values that can be adjusted according to the data collection setups (controlled vs in-the-wild, remote vs near-eye cameras... etc) and the corresponding established error values (e.g. eye-tracker drift or calibration errors).

\section{Experiments}
\label{sec:experiments}
\subsection{Datasets}

To evaluate \methodName, we use different appearance-based gaze estimation datasets covering different conditions: appearances (genders, ethnicities, glasses, and make-up), illumination (indoor and outdoor), and gaze direction and head pose distributions. Mainly, experiments were conducted on the full-face MPIIFaceGaze \cite{zhang2017s} and GazeCapture \cite{krafka2016eye} datasets.
The MPIIFaceGaze \cite{zhang2017s} dataset contains $\sim$200 thousand full-face images collected in the wild from 15 participants. The GazeCapture \cite{krafka2016eye} dataset contains $\sim$2.5 million frames of $\sim$1.5 thousand participants.

\subsection{Results and Implementation Details}

We train our VAE with an enhanced ResNet \cite{wang2023investigation} backbone on the training set (80\% of GazeCapture) in a person-specific fashion since the inter-subject anatomical differences are known to affect the performance of gaze-based tasks \cite{krafka2016eye, zhang2017s}. We then test our model on the remaining 20\% of GazeCapture and the full MPIIFaceGaze dataset.

\subsubsection{\textbf{Training the VAE}}
We use the well-established loss function \cite{park2019few} adapted to our disentanglement criteria: 
$$L_{\text{full}} =  \lambda_{recon} L_{\text{recon}} + \lambda_{EC} L_{\text{EC}} + \lambda_{gaze} L_{\text{gaze}} + \lambda_{KL}L_{\text{KL}}$$
where $L_{\text{recon}}$ is the reconstruction loss that guides the encoding-decoding process pixel-wise, $L_{\text{EC}}$ is the embedding consistency loss that ensures the embedding of the same appearance into the same features even with different (disentangled) gaze direction and head pose, $L_{\text{gaze}}$ is the gaze direction loss between the estimated gaze of the MLP and the true gaze direction, and $L_{\text{KL}}$ is the VAE Kullback-Leibler divergence loss that regularizes the model by approximating the prior distribution via the encoded distribution and penalising deviations from the model. 
For the coefficients, we use $\lambda_{recon} = 2$, $\lambda_{EC} = 1$, $\lambda_{gaze} = 0.1$, and $\lambda_{KL} = 1$ with a batch size of $128$ and a learning rate of $ 5\cdot10^{-7}$.
This yielded a reconstruction loss of $0.3859$, a gaze angular error of $5.0543^\circ$, and a KL-divergence loss of $12.9531$. The $EC_{gaze}$ is $5.8974$ and the $EC_{pose}$ is $10.4678$.

\subsubsection{\textbf{Feature Disentanglement}} 
\begin{figure}
\centering
\begin{minipage}{.45\textwidth}
  \centering \includegraphics[width=.95\linewidth]{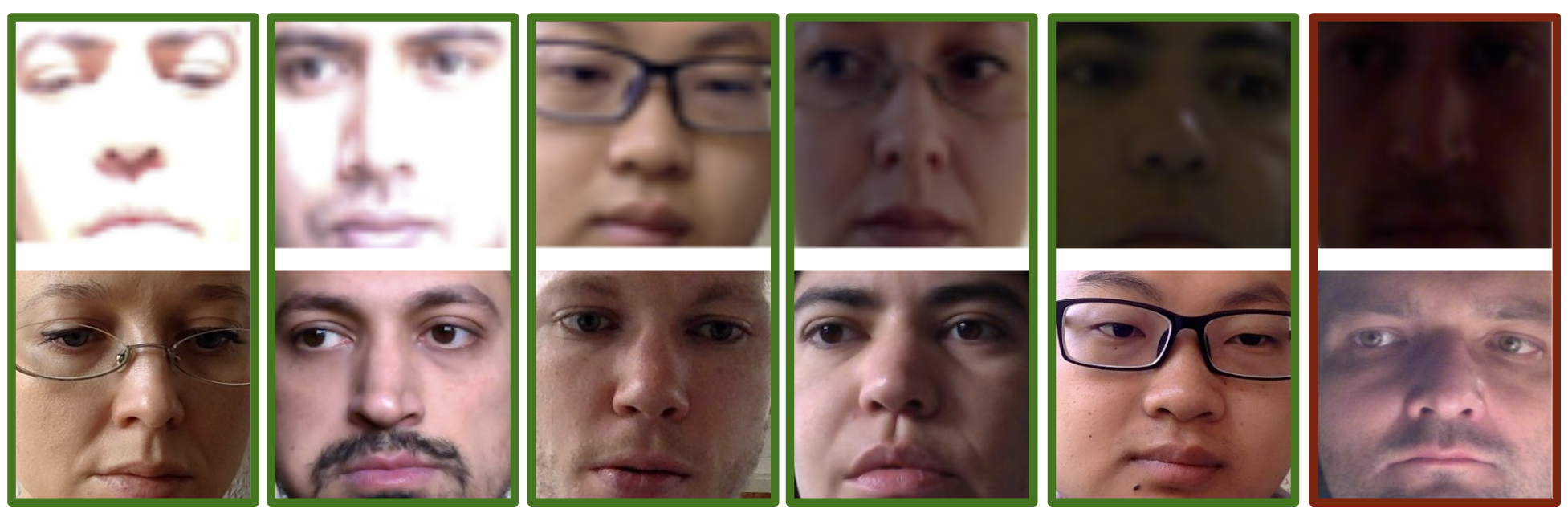}
  \captionof{figure}{A qualitative example of the correct predictions (green) and incorrect predictions (red) of the gaze direction verification. Each pair represents different aspects of appearance, e.g. different subjects, head poses, lighting conditions, glasses, makeup, genders, and race. Therefore, QualitEye can successfully disentangle the appearance code, removing its effect from the overall method.}
  \label{fig:appearance}
\end{minipage}%
\hspace{0.05\textwidth}
\begin{minipage}{.45\textwidth}
  \centering
  \includegraphics[width=.95\linewidth]{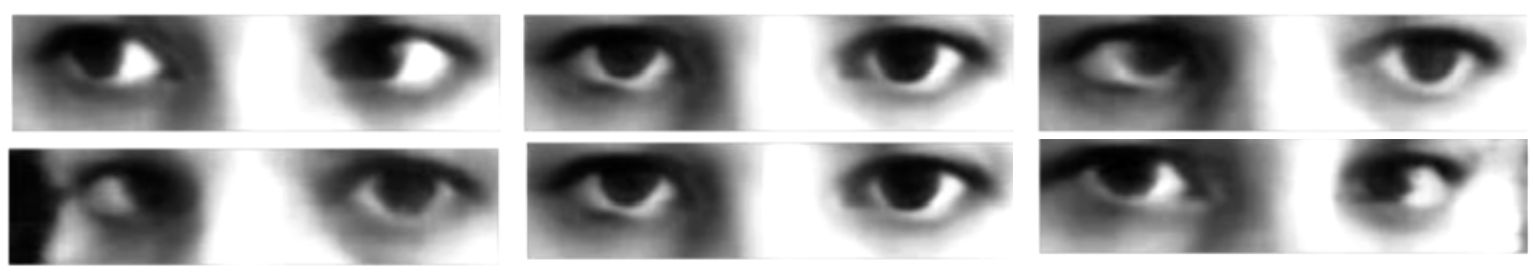}
  \captionof{figure}{A qualitative example of the output of the decoder after the disentanglement and rotation of gaze direction (top, with a fixed head pose and rotating gaze direction) and head pose (bottom, with a fixed gaze direction and rotating head pose).}
  \label{fig:rotation}
\end{minipage}
\end{figure}

\begin{figure}
\centering
\begin{minipage}{.34\textwidth}
  \centering \includegraphics[width=.9\linewidth]{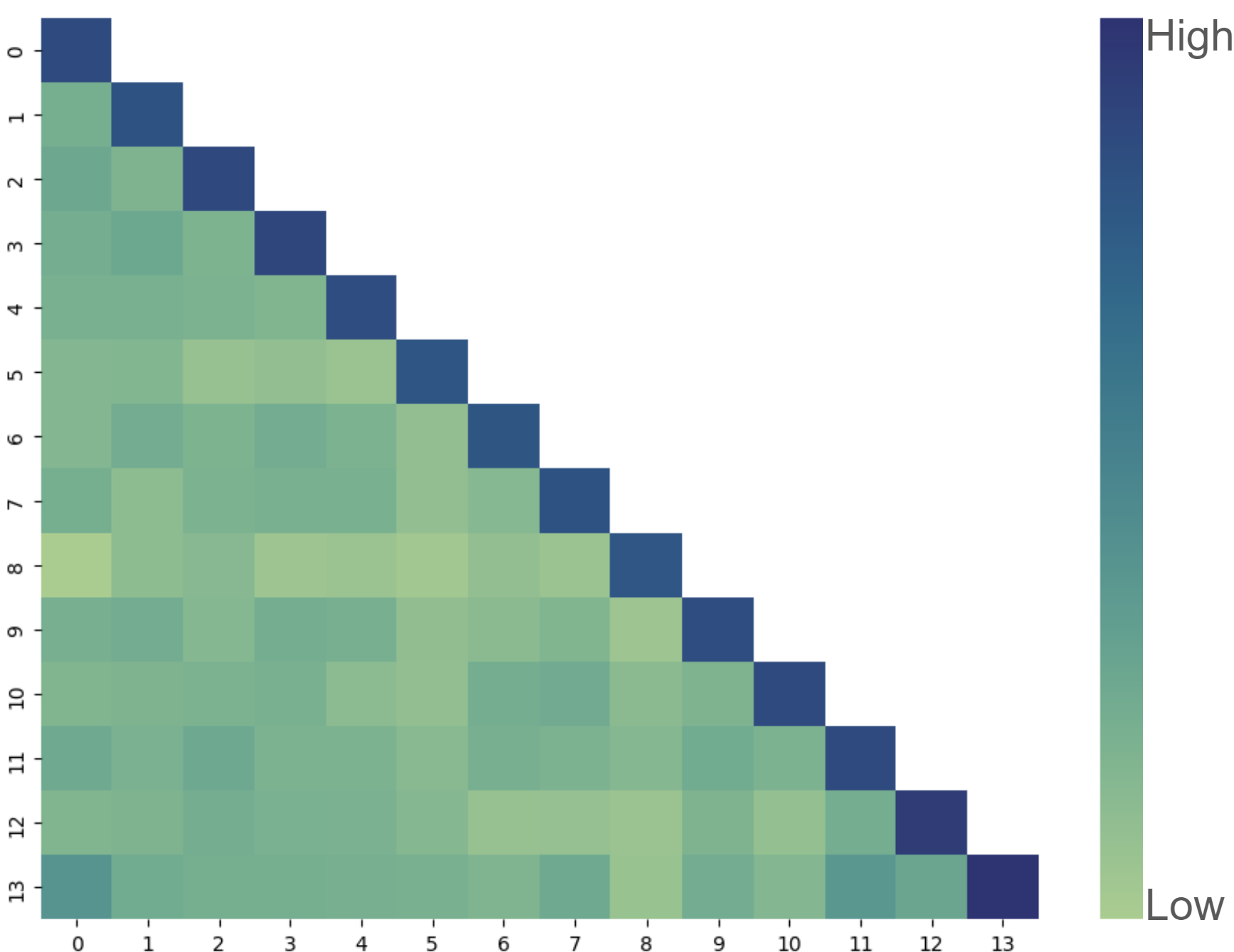}
  \captionof{figure}{Cross-participant performance as normalised MCC on the \\ MPIIFaceGaze dataset}
  \label{fig:MCC_person}
\end{minipage}%
\hspace{0.05\textwidth}
\begin{minipage}{.6\textwidth}
  \centering
  \includegraphics[width=.75\linewidth]{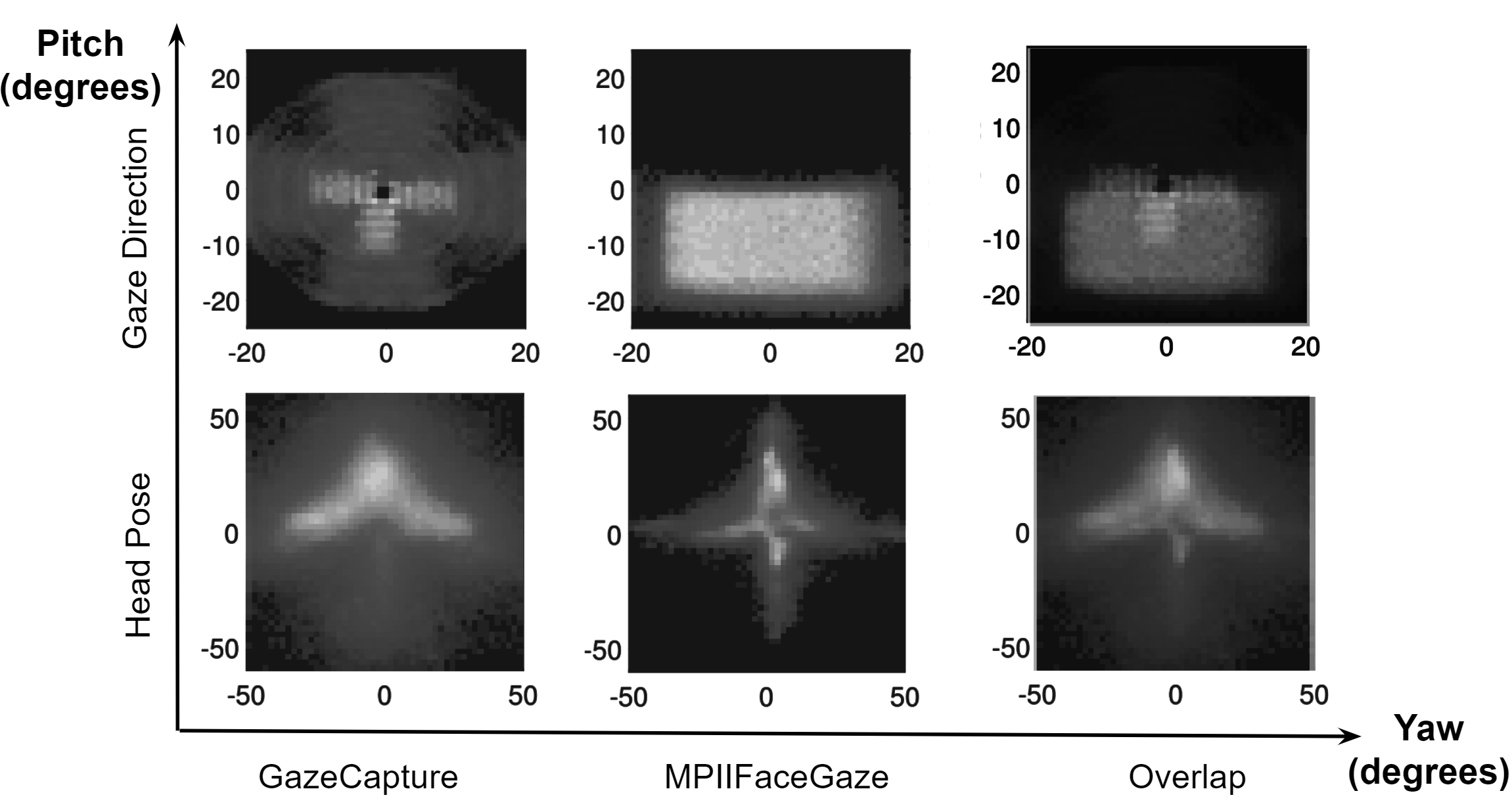}
  \captionof{figure}{The pitch and yaw distribution of gaze direction and head pose on the GazeCapture (reference $R$) dataset, the MPIIFaceGaze (owner $O$) dataset, and the corresponding distribution overlap. QualitEye captures the overlapping samples and misses the samples in $O$ that do not have a matching sample in $R$.}
  \label{fig:dist}
\end{minipage}
\end{figure}

As shown in Figure \ref{fig:VAE}, to disentangle the appearance, gaze direction, and head pose features, the encoder processes the input image into three distinct latent codes of size 64, 2, and 16, respectively.
\\
Qualitatively, as shown in Figure \ref{fig:appearance}, we intentionally neglect the appearance code for (i) data minimisation, i.e. sharing less information to enhance privacy and (ii) for transferability across different subjects. The head pose code captures rotation and orientation (i.e. pitch and yaw angles) and excludes gaze direction to ensure that the gaze features remain consistent even if head orientation changes. 
This is achieved by a transformation (i.e. a rotation matrix), proposed by Park et al. \cite{park2019few}, that maps each gaze representation relative to a canonical (or frontal) head position, leading to successfully disentangling the gaze direction and head pose as shown in Figure \ref{fig:rotation}. 
\\
Quantitatively, correctly matching samples between source and reference datasets provides a direct quantitative indicator of successful feature disentanglement.
We report the quality metric as a Matthews correlation coefficient (MCC) \cite{matthews1975comparison}, to account for true positives (TP, the correct compliant matching samples), true negatives (TN, the correct non-compliant mismatching samples), false positives, and false negatives (FP and FN, the incorrect predictions).
MCC is particularly interesting as it can be used for classes of different sizes \cite{boughorbel2017optimal}, i.e. both symmetric and asymmetric scenarios.
A coefficient of $+1$ indicates a perfect match, $0$ represents a random prediction, and $-1$ is a total disagreement between the predicted match and the true match.
\\
We run experiments on the within- and cross-participant for the same domain, e.g. MPIIFaceGaze in \autoref{fig:MCC_person}. We further report the average MCC with varying participant-based data splits to handle the unbalanced data distribution across data sources in \autoref{tab:MCC_participant}, and cross-datasets in \autoref{tab:MCC_cross} quality verification. Note that, we use the available dataset's labels as our ground-truth which is subject to error since creating labels is a complex task in the first place given the eye-head interplay, eye registration error, occlusions, appearance biases... etc. Hence, we adapt our hashing step to compensate for such (unavoidable) errors in the data where the dimensionality of the target projected space is reduced to 80 bits with a collision probability of 0.05\footnote{Values are calculated according to the current SOTA remote gaze estimation models \cite{wang2023investigation}. For instance, the normal gaze range is [-45,45] with a few outliers and an error of 3 and 5 degrees for gaze direction and head pose, respectively.}.

Hence, the disentanglement is not achieved implicitly (e.g., via adversarial losses), but rather by explicitly enforcing how specific latent variables must behave under known physical transformations (i.e., geometric rotations).

\begin{table}[t]
    \centering
    \caption{Within-dataset performance for gaze verification averaged over different participant-based data splits.}
    \begin{tabularx}{1\textwidth} 
    { 
       >{\raggedright\arraybackslash}X 
       >{\raggedright\arraybackslash}X
       >{\centering\arraybackslash}X
       >{\centering\arraybackslash}X
       >{\centering\arraybackslash}X
       >{\centering\arraybackslash}X
       >{\centering\arraybackslash}X
      }
     \toprule
      & Dataset & TP & TN & FP & FN & \mbox{\textbf{MCC}}  \\
     \midrule
     \textbf{Gaze} & MPIIFaceGaze & $0.9628$ & $0.9967$ & $0.0033$ & $0.0372$ & $0.9220$  \\
     \textbf{Direction} & GazeCapture &   $0.9808$ & $0.9966$ & $0.0034$ & $0.0192$ & $0.9402$ \\
     \midrule
     \textbf{Head Pose} & MPIIFaceGaze & $0.8381$ & $0.9062$ & $0.0938$ & $0.1619$ & $0.746$  \\
     & GazeCapture &   $0.9564$ & $0.9289$ & $0.0711$ & $0.0436$ & $0.8856$ \\
     \bottomrule
    \end{tabularx}

    \label{tab:MCC_participant}
\end{table}

\begin{table}[t]
    \centering
    \caption{Cross-domain performance for gaze quality verification on a TITAN X 12G GPU}
\begin{tabularx}{\textwidth}{
  >{\raggedright\arraybackslash\hsize=2\hsize}X
  >{\centering\arraybackslash\hsize=0.8\hsize}X
  >{\centering\arraybackslash\hsize=0.8\hsize}X
  >{\centering\arraybackslash\hsize=0.8\hsize}X
  >{\centering\arraybackslash\hsize=0.8\hsize}X
  >{\centering\arraybackslash\hsize=0.8\hsize}X
  >{\centering\arraybackslash\hsize=0.8\hsize}X
}
     \toprule
      Dataset & TP & TN & FP & FN & \mbox{\textbf{MCC}} & Runtime \\
     \midrule
     MPIIFaceGaze/ GazeCapture & $0.9740$ & $0.9966$ & $0.0360$ & $0.0260$ & $0.9331$ & $152s$\\
     \bottomrule
    \end{tabularx}
    \label{tab:MCC_cross}
\end{table}

\subsubsection{\textbf{Datasets' Discrepencies}}
Although the GazeCapture and MPIIFaceGaze datasets are widely used in gaze research, they differ significantly in terms of data collection methodology, device settings, environmental conditions, and participant demographics.
As shown in Table \ref{tab:MCC_participant} and Figure \ref{fig:MCC_person}, our method was able to successfully disentangle gaze direction and head poses within the same dataset, regardless of the appearance differences such as the different participants (1400 vs 15), demographics (diverse range of age vs university students), and background and lightning conditions (outdoor vs indoor) in GazeCapture and MPIIFaceGaze, respectively. The performance slightly degrades for head pose since the GazeCapture dataset was self-collected by participants using mobile devices (iOS phones and tablets), leading to a high variation in head pose due to uncontrolled conditions. It was also annotated with gaze points that are less precise due to mobile device limitations. Meanwhile, the MPIIFaceGaze dataset was recorded using laptops with a front-facing camera in indoor setups. This led to limited head pose variation and precise annotations due to controlled lab settings. Hence, the GazeCapture dataset is larger, more variable, and more general, and the same domain was used to train the VAE, hence the better performance.
\newline
Nonetheless, QualitEye mainly relies on a reference dataset for comparison. Therefore, as shown in Table \ref{tab:MCC_cross} and Figure \ref{fig:dist}, \methodName~ also succeeds at comparing different domains (i.e. datasets) but mainly fails in cases where a sample in one dataset does not have a matching sample in the other dataset when comparing ground-truth labels (i.e. outliers). In practice, carefully selecting a reference dataset that meets all requirements is recommended.

\subsubsection{\textbf{Privacy-preserving gaze data quality verification}}

The private cross-party gaze data verification setups maintain the same performance as the local public one in terms of data quality metrics. We use a computational security parameter $k$ = 128 and a statistical security parameter of $\sigma$ = 40 following \cite{kolesnikov2016efficient}.\footnote{
$k$ = 128, $\sigma$ = 40 is a standard security parameter choice. Other values are possible. Generally, lower values might reduce the runtime of the cryptographic parts but can be more vulnerable to attacks. The security parameter choice does not affect accuracy.}
However, privacy comes with a computational overhead in runtime and communication. Since the magnitude of this overhead depends on multiple factors, e.g., dataset size (c.f. Section \ref{sec:private}), our proposed versions account for the practical runtime while maintaining the desired privacy guarantees, as shown in \autoref{tab:runtime}.

\begin{table}[H]
    \centering
    \caption{LAN online runtime in ms as overhead with respect to the public verification runtime (\methodName$_{v0}$) on a TITAN X 12G GPU for different dataset sizes: $2^8$ (small-hundreds),  $2^{12}$ (small-thousands), $2^{18}$ ($\sim$ MPIIFaceGaze), $2^{21}$ ($\sim$ GazeCapture), and $2^{24}$ (larger datasets). The table shows the symmetric scenarios, however different asymmetric splits are also possible (c.f. \ref{tab:MCC_cross}).}
    \begin{tabularx}{1\textwidth} 
    { >{\raggedright\arraybackslash}X 
       >{\centering\arraybackslash}X
       >{\centering\arraybackslash}X
       >{\centering\arraybackslash}X 
       >{\centering\arraybackslash}X
       >{\centering\arraybackslash}X
      }
     \toprule
      
     Protocol/Size & $2^8$ & $2^{12}$ & $2^{18}$ &   $2^{21}$ & $2^{24}$\\
     \midrule
     \methodName$_{v0}$ & 128 & 2,048 & 131,072 & 1,048,576 & 8,388,608 \\
     \midrule
     \methodName$_{v1}$ & $+ \sim 100$ & $+ \sim$ 1,600 & $+ \sim$ 102,400 & $+ \sim$ 819,200 & $+ \sim$ 6,553,600 \\
     \methodName$_{v2}$ & $+ \sim 101$ & $+ \sim$ 1,601 & $+ \sim$ 102,402 & $+ \sim$ 819,202 & $+ \sim$ 6,553,602 \\
     \methodName$_{v3}$ & $+ \sim 51$ & $+ \sim 802$ & $+ \sim$ 51,202 & $+ \sim$ 409,602 & $+ \sim$ 3,276,802 \\
     \methodName$_{v4}$ & $+ \sim 201$ & $+ \sim 223$ & $+ \sim$ 1,284 & $+ \sim$ 13,894 & $+ \sim$ 58,600\\
     \bottomrule
    \end{tabularx}

    \label{tab:runtime}
\end{table}

\subsubsection{\textbf{Robustness vs. Privacy}}

We assess robustness by comparing hashes of the disentangled features (Table \ref{tab:MCC_participant}) to the original images (without disentanglement). 
For the latter, we also modify the appearance, while maintaining the gaze and pose vectors, with different collision and evasion attacks (c.f. Appendix \ref{apx:attacks}):
Collisions correspond to incorrectly matching samples with different gaze or head pose, while evasion attacks correspond to failing to match samples with the same gaze/pose due to minor appearance changes. Our results show that hashing raw images is highly vulnerable to evasion (small appearance changes cause large hash differences - MCC = -0.87 and -0.9 for MPIIGaze and GazeCapture), whereas disentangled perceptual hashes strike the desired balance: They resist evasion via irrelevant appearance changes while maintaining low collision rates across genuinely different gaze configurations.
\\
We also evaluate privacy. Perceptual hashes are irreversible, as they compress the input into a low-dimensional code that preserves only the features relevant for similarity comparison while discarding all other details. By comparing images of the same participant (i.e., same appearance), we find that the similarity metric remains too low to reliably map them to the same person (MCC = 0.01 and -0.07 for MPIIGaze and GazeCapture, respectively), indicating that appearance information is effectively removed.
Beyond appearance, no other attributes can be reliably inferred. Although the datasets utilised in this study do not provide participant demographic information, precluding any further analysis based on age, gender, ethnicity, or other subject-specific factors, the irreversibility of the mapping from an input $x$ to a disentangled latent representation $z = E(x)$ and subsequently to a perceptual hash $s = h(z)$ can be formalised using information-theoretic arguments. 
The perceptual hash $h(\cdot)$ compresses $z$ into a fixed-length code $s \in \{0,1\}^d$. Because the latent space $\mathcal{Z}$ has much higher intrinsic dimensionality than the hash length $d$, the composition $F = h \circ E: \mathcal{X} \to \{0,1\}^d$ is many-to-one, i.e., multiple distinct inputs $x_1 \neq x_2$ may yield the same hash $s$. 
From an information-theoretic perspective, the entropy of the input satisfies $H(X) \gg H(S) \le d$, implying that the mutual information $I(X;S) \le H(S) \ll H(X)$, and therefore no deterministic or probabilistic function can reconstruct $x$ from $s$.
Together, robustness to irrelevant variations and irreversibility of sensitive details make perceptual hashes essential for \methodName.

\section{Discussion}

Our results show that the gaze data quality verification problem can be solved efficiently under different public and privacy-preserving setups while achieving a good tradeoff between performance, runtime, and communication. 
\\
Our experiments demonstrate that the proposed VAE effectively disentangles appearance, gaze direction, and head pose into separate latent codes. 
Gaze and head pose representations remain consistent across variations in appearance (MCC gaze: 0.92–0.94, head pose: 0.75–0.89). This success is achieved by explicitly enforcing known physical transformations, i.e. geometric rotations, rather than relying on implicit adversarial objectives. 
\methodName~ also performs robustly across datasets with different characteristics (MCC = 0.93), demonstrating that the method can handle discrepancies in environment, devices, and annotation quality. 
In addition, the privacy-preserving implementation maintains the same verification performance as the public setup while introducing manageable runtime overhead. 
The hashes are irreversible, preventing the inference of other participant attributes beyond gaze and head pose. 
We further confirm our theoretical hypotheses in \autoref{sec:method}: \methodName$_{v1}$ can be used when both parties have relatively small datasets (i.e. tens to hundreds of samples) and would like to know the exact (mis)matching samples. \methodName$_{v2}$ can be used when both parties have relatively small datasets (i.e. tens to hundreds of samples) and are only interested in knowing the cardinality of the intersection with an additional shuffling step. \methodName$_{v3}$ can be used when one party has a relatively small dataset while the reference party owns a larger dataset (i.e. thousands to billions of samples) by offloading the larger dataset computation to an offline phase, i.e. asymmetric scenarios. \methodName$_{v4}$ can be used when one or both parties own large datasets as the runtime with OPRF-based approach decreases significantly with respect to the DH-based approaches when the dataset size increases.

\paragraph{\textbf{Limitations and future work}}
We focus on gaze angles and head poses, as they are the most commonly available labels in gaze-based datasets. While this problem has not been previously studied, we hypothesise that gaze data quality verification can be further improved by disentangling additional factors, such as illumination conditions \cite{jiang2023nerffacelighting}. We restrict our scope to image-based gaze tasks, noting that other modalities (e.g., scanpaths or videos) require specialised models and privacy assumptions due to temporal dependencies. Although we present a two-party computation (2PC) protocol, it can be extended to multi-party computation (MPC), where incorporating additional data sources may further reduce verification errors. Finally, \methodName~assumes a semi-honest threat model\footnote{The semi-honest model aligns with our intended deployment in collaborative research and data-sharing settings where parties (e.g., research groups or dataset providers) are expected to follow the protocol but may attempt to infer additional information from exchanged messages, without manipulation.}; stronger security guarantees under malicious adversaries are possible, eliminating collision and evasion attacks, albeit reduced efficiency \cite{pinkas2020psi}.

\section{Conclusion}

We presented \methodName-- the first work to investigate the problem of gaze data quality verification. We introduced a new generic hashed representation learning model that disentangles the gaze direction and head pose features for a high-domain adaptation performance, ignoring the cross-user irrelevant features (e.g. appearance) to allow for a label-specific comparison.
Furthermore, we extended existing privacy-preserving interactive protocols with semantic similarities and labels matching to handle the different privacy and trust requirements. Our results show that \methodName~ is efficient under different public and privacy-preserving setups in terms of performance, runtime, and communication.

\section*{Privacy and Ethics Statement}
\methodName~ is designed with privacy preservation as a core principle. 
It enables broader data sharing and cross-domain evaluation without requiring access to raw eye images, thereby supporting reproducibility and collaboration while respecting user privacy.
This can benefit applications in human-computer interaction, accessibility, and healthcare, where high-quality gaze data is essential, but privacy concerns are paramount.
We, therefore, emphasise that any deployment of the proposed method should adhere to established data protection regulations (e.g., GDPR), ensure informed user consent, and incorporate transparency about how gaze data is processed and shared.
All datasets used in this work are publicly available. To the best of our knowledge, there are no harmful applications that can be implemented on top of \methodName, and its design further mitigates potential misuse.

% ...

\bibliographystyle{ACM-Reference-Format}
\bibliography{references}

\appendix
\section{Collision and Evasion Attacks}\label{apx:attacks}

In the context of hashing-based and similarity-driven verification, adversarial behaviour is commonly characterised by \emph{collision} and \emph{evasion} attacks, which target different failure modes of the representation.
We build our attacks on the work of Struppek et al \cite{struppek2022learning}. Struppek et al. analyse the robustness of perceptual image hashing by explicitly formulating two complementary attack models: collision attacks and evasion attacks, both realised as optimisation problems over the image space.

\subsection{Collision Attacks} A collision attack aims to identify two distinct inputs $x_1 \neq x_2$ that produce the same hash value, i.e.,
$h(x_1) = h(x_2)$.
The objective is to falsely link or impersonate samples that are semantically different. In perceptual or similarity hashing, collisions are inherently possible due to the low-dimensional and many-to-one nature of the hash space. However, a secure and task-appropriate hash function ensures that collisions occur only for inputs that are semantically equivalent with respect to the target attributes (e.g., identical gaze direction and head pose), while remaining unlikely for dissimilar inputs. \\
Technically, an adversary starts with a target hash (or target image) and iteratively modifies another image so that its perceptual features align with the same low-dimensional representation used by the hash function. Because perceptual hashes rely on compressed, robust features, attackers manipulate precisely those components while allowing high-frequency or imperceptible changes elsewhere. 
This manipulation is done through a gradient-based collision attack. Since the final binary hash is non-differentiable, the attack operates on the continuous internal embedding produced by the hashing model before thresholding. An optimization objective is defined that (i) encourages the embedding of the modified image to align with the target embedding and (ii) penalizes perceptual distortion using similarity metrics. Using gradient-based optimization methods (e.g., projected gradient descent or Adam), the image pixels are iteratively adjusted in the direction that reduces hash distance, with updates constrained to valid pixel ranges. The attack succeeds because small changes in the continuous embedding—especially near threshold boundaries—can flip multiple hash bits, allowing two semantically different images to produce identical or near-identical perceptual hashes.
The result is two semantically different images that are deemed similar by the hashing system, undermining content authentication or deduplication.

\subsection{Evasion Attacks}
An evasion attack seeks to modify an input $x$ into $x'$ such that
$h(x') \neq h(x)$,
despite the modification being visually imperceptible or semantically irrelevant. The goal is to prevent correct matching by exploiting the excessive sensitivity of the hashing function to nuisance factors such as lighting, texture, or minor appearance variations. Evasion attacks, therefore, directly assess the robustness of the representation to irrelevant perturbations.

In practice, the attacker starts from an original image and iteratively perturbs it to increase the Hamming distance between the original hash and the modified hash beyond the detection threshold. Because the final hash is binary and non-differentiable, the attack operates on the continuous internal embedding produced by the model before binarization. By computing gradients of a loss that encourages divergence in embedding space, the attacker identifies which pixels most strongly influence specific hash bits.
The optimization is constrained by a perceptual similarity metric—such as MSE, SSIM, or LPIPS—to ensure that the modified image remains visually close to the original. At each iteration, the algorithm adjusts pixel values in directions that maximally shift the embedding while keeping distortion within a predefined bound. Since perceptual hashes typically rely on thresholding continuous features (e.g., sign of embedding components), small shifts near decision boundaries can flip multiple bits simultaneously. The attack therefore focuses on regions or frequency components that disproportionately affect these embedding dimensions.

\subsection{Key Distinctions}
The key distinction is that collision attacks cause \emph{false matches} between different samples, whereas evasion attacks cause \emph{missed matches} between equivalent samples.

In this work, we (i) fine-tune and optimise the models to our domain and, in addition to \cite{struppek2022learning}, (ii) incorporate mean squared error (MSE) and Learned Perceptual Image Patch Similarity (LPIPS). 
While SSIM captures structural and luminance-based similarities aligned with human perception, MSE provides a pixel-wise measure of absolute intensity differences, and LPIPS evaluates perceptual similarity in a deep feature space learned by neural networks. 
Using these complementary metrics allows us to assess robustness across low-level, structural, and semantic perturbations, yielding a more comprehensive characterisation of image modifications and their impact on perceptual hashing.

\begin{figure}
        \centering
        \begin{subfigure}{0.9\textwidth}
            \centering
            \includegraphics[width=0.8\textwidth]{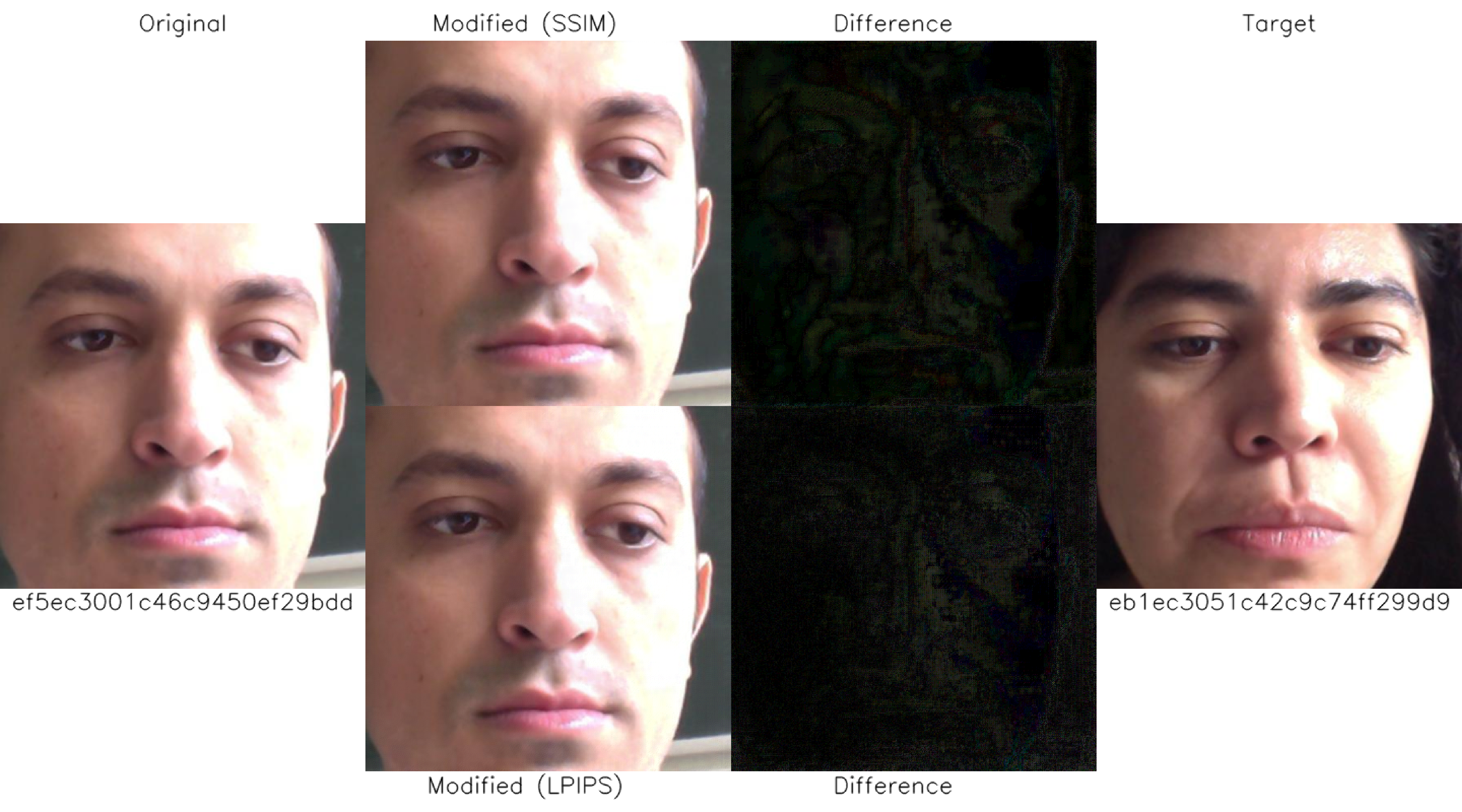}
        \end{subfigure}
        \begin{subfigure}{0.9\textwidth}
            \centering
            \includegraphics[width=0.8\textwidth]{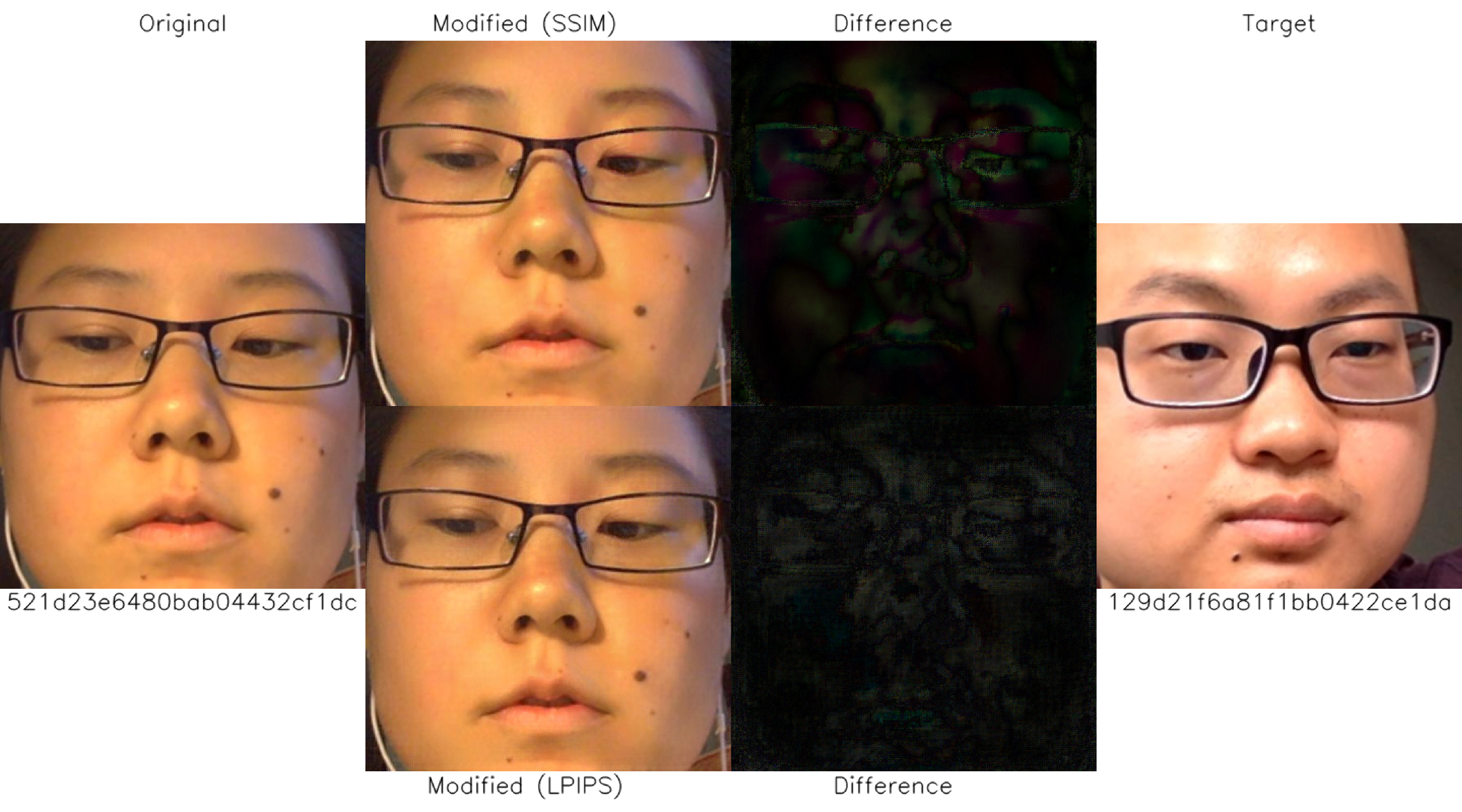}
        \end{subfigure}
        \begin{subfigure}{0.9\textwidth}
            \centering
            \includegraphics[width=0.8\textwidth]{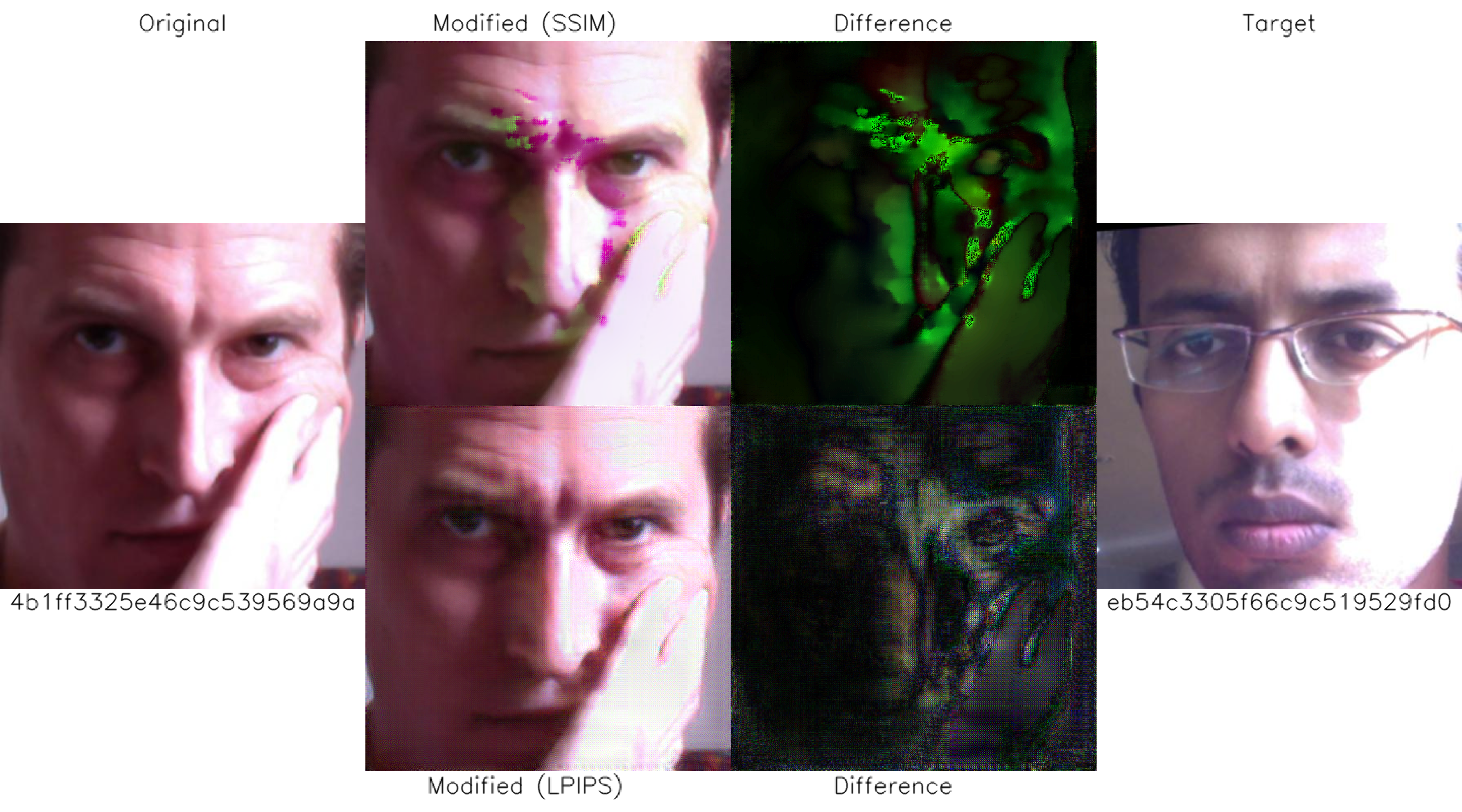}
            \vspace{-1em}
        \end{subfigure}
        \caption{Collision attack results. Each row presents the original image (left), the adversarially modified image (middle), and the target image (right) for a different participant. The corresponding difference plots for the modified images are also displayed; these plots are generated by subtracting the original and modified images using the absolute difference and scaling the result to enhance the visibility of changes. Below the original image, the starting hash value is displayed in hexadecimal, while below the target image, the hash value of the modified image is shown.}
        \label{fig:collision}
\end{figure}

\begin{figure}[p] 
\raggedright
\rotatebox{90}{%
  \begin{minipage}{1\textwidth} 
    \centering
    \includegraphics[width=2\textwidth,height=\textwidth,keepaspectratio]{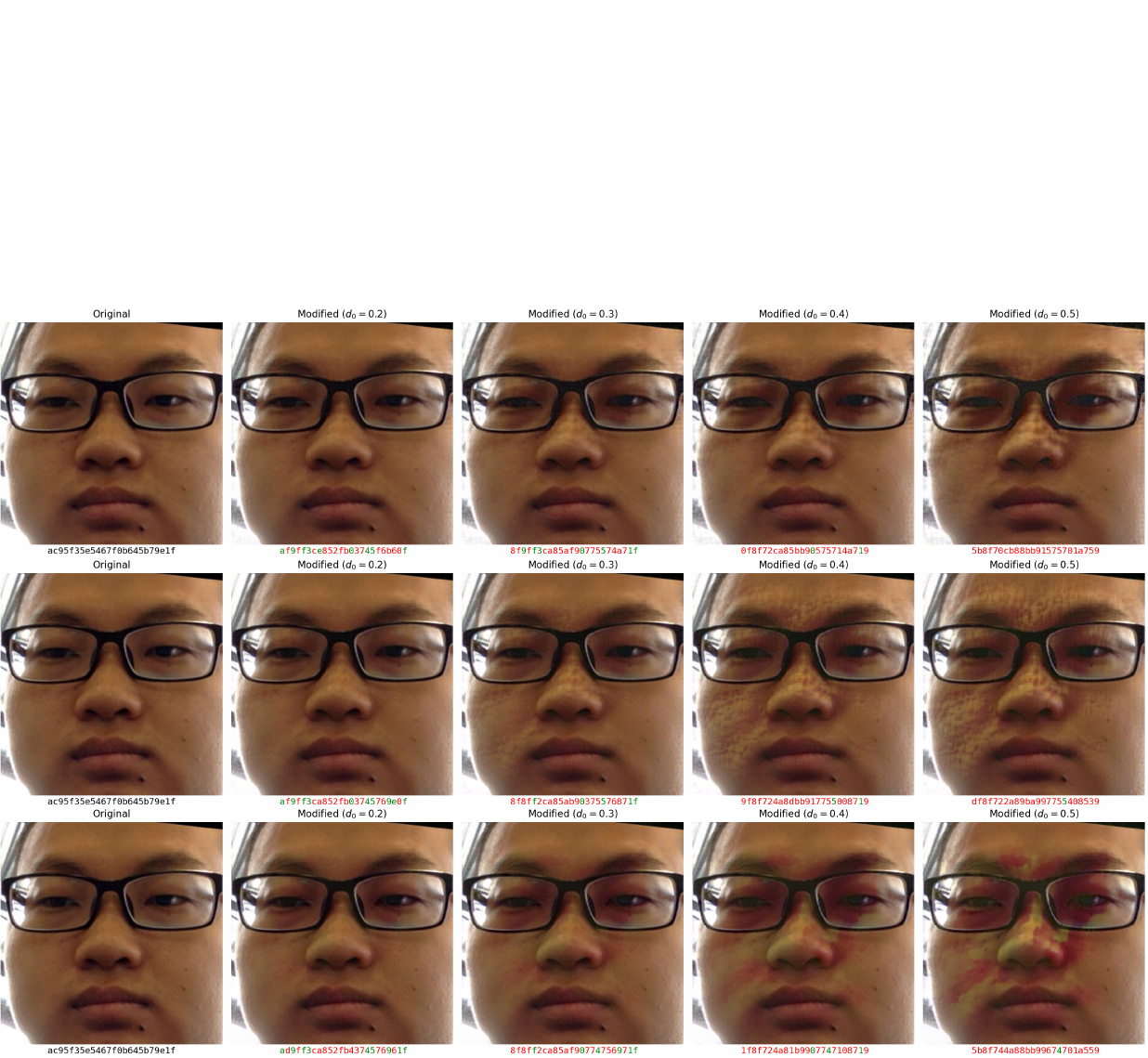} 
    \caption{Evasion modification with $LPIPS$ (top), $MSE$ (middle), and $SSIM$ (bottom) with thresholds ($0.2$ to $0.5$) with evasion attacks. Hash values are displayed below each image in hexadecimal format, with modified hashes colour-coded: red for changes and green for no change.}
    \label{fig:evasion}
  \end{minipage}%
}
\end{figure}

Our qualitative results are shown in Figures \ref{fig:collision} and \ref{fig:evasion}. When passing the resulted images to our pipeline, as discussed in the main text, perceptual hashes computed from full images without disentanglement are highly susceptible to attacks, whereas hashes derived from disentangled features exhibit strong robustness and are largely resistant to both collision and evasion attempts.

\end{document}